%% file: aprile.tex
\documentclass[graybox]{svmult}

\usepackage{mathptmx}       
\usepackage{helvet}         
\usepackage{courier}        
\usepackage{type1cm}        
\usepackage{makeidx}         
\usepackage{graphicx}        
\usepackage{multicol}        
\usepackage[bottom]{footmisc}
\usepackage{lineno}

\makeindex             

\newcommand{\Xehund}{{\sc Xenon100}}
\newcommand{\Xe}{{\sc Xenon}}
\newcommand{\Xeton}{{\sc Xenon1T}}
\newcommand{\Xeten}{{\sc Xenon10}}
\newcommand{\n}[1]{\mathrm{#1}}

\begin{document}

\title*{The \Xeton\ Dark Matter Search Experiment}
\author{E. Aprile, Columbia University, on behalf of the \Xe\ collaboration}
\institute{Elena Aprile, Columbia University, New York, NY 10027, USA. \email{age@astro.columbia.edu}}
%
%
\maketitle

\abstract{The worldwide race towards direct dark matter detection in the form of Weakly Interacting Massive Particles (WIMPs) has been dramatically accelerated by the remarkable progress and evolution of liquid xenon time projection chambers (LXeTPCs). With a realistic discovery potential, \Xehund\ has already reached a sensitivity of $7\times10^{-45}\,\n{cm}^2$, and continues to accrue data at the Laboratori Nazionali del Gran Sasso (LNGS) in Italy towards its ultimate sensitivity reach at the $\sigma_{\n{SI}}\sim 2\times10^{-45}\,\n{cm}^2$ level for the spin-independent WIMP-nucleon cross-section. 
To fully explore the favoured parameter space for WIMP dark matter in search of a first robust and statistically significant discovery, or to confirm any hint of a signal from \Xehund, the next phase of the \Xe\ program will be a detector at the ton scale - \Xeton. 
The \Xeton\ detector, based on 2.2~ton of LXe viewed by low radioactivity photomultiplier tubes and housed in a water Cherenkov muon veto at LNGS, is presented.  
With an experimental aim of probing WIMP interaction cross-sections above of order $\sigma_{\n{SI}}\sim 2\times10^{-47}\,\n{cm}^2$ within 2 years of operation, \Xeton\ will provide the sensitivity to probe a particularly favourable region of electroweak physics on a timescale compatible with complementary ground and satellite based indirect searches and with accelerator dark matter searches at the LHC. Indeed, for a $\sigma_{\n{SI}} \sim 10^{-45}\,\n{cm}^2$ and $100~\,\n{GeV/c^2}$ WIMP mass, \Xeton\ could detect of order 100 events in this exposure, providing statistics for placing significant constraints on the WIMP mass.}

\section{The \Xeton\ Detector}
\label{sec:xeton}

Building upon the success of the \Xe\ detectors thus far, we will develop and deploy the next generation of detector in the program - \Xeton. 
The \Xeton\ instrument can be realized by essentially scaling up the existing \Xehund\ detector by about a factor of 10 and reducing the background by a factor of 100. This was successfully achieved in going from \Xeten~\cite{Angle:2008} to \Xehund~\cite{Aprile:2011instr, Aprile:2011Run08}. Employing increased levels of self-shielding and identical techniques for particle discrimination, the technologies required are largely already proven, and ongoing R$\&$D efforts address those that are not immediately transferable from \Xehund.  
The \Xeton\ detector is a dual-phase TPC containing 2.2~ton of pure LXe instrumented with PMTs for the simultaneous detection of  scintillation light and ionization charge via proportional gas scintillation. The PMTs (Hamamatsu R11410 series) have a maximum sensitivity  at the peak of the Xe scintillation emission spectrum (178 nm). Approximately 250 PMTs, 3-inch in diameter, are arranged in two closely-packed arrays: a ``bottom'' array with $\sim$120 mounted in the liquid, below the TPC drift volume, and a ``top'' array with $\sim$130 mounted in the gas above the liquid. The approximate 1.1~ton active volume is defined by a 1~m diameter cylinder that is also 1~m high, made out of PTFE (teflon) for its high reflectivity in the VUV region. Wire meshes with high optical transmission close the cylinder, and define the LXe drift volume and the gas Xe proportional region, above the liquid level. Field shaping electrodes made of copper are mounted on the outside of the PTFE cylinder and define a homogenous electric drift field (1~kV/cm) within the active volume. As in \Xeten\ and \Xehund, the Xe gas is liquefied and kept at the desired temperature by PTRs, coupled directly to the inner volume.  

The \Xeton\ experiment will be mounted in Hall B at LNGS, between the ICARUS and WArP experiments.  The infrastructure consists of two main elements: a 9.6~m diameter water tank as shield and active Cherenkov muon veto with 4$\pi$ coverage for the \Xeton\ detector, and a service building that contains the cryogenic infrastructures and purification systems,  the Xe storage/recovery system and the DAQ and control electronics. These structures will be in place for the entire duration of the project, while a mobile platform and clean room will be available during the initial assembly phase in the water shield and during maintenance operations. 

Detailed simulation studies informed by results from previous \Xe\ and other LXe detectors indicate that an increase in the light yield of \Xeton\ relative to \Xehund\ is achievable by adopting relatively modest modifications to the design of the TPC such as greater coverage of non-reflective surfaces with PTFE of near unity reflectivity, greater optical transmission of electrode structures and especially greater quantum efficiency (QE) of photomultiplier tubes (PMTs) at the 178~nm wavelength of Xe scintillation. At 5~m absorption length, the predicted average light yield at 122~keV$_{\n{ee}}$ is 3~photoelectrons/keV$_{\n{ee}}$ at the nominal operating field of 1~kV/cm, corresponding to a nuclear recoil threshold below that of \Xehund. 

\begin{figure}[t]
\centering
\includegraphics[width = 1.\columnwidth]{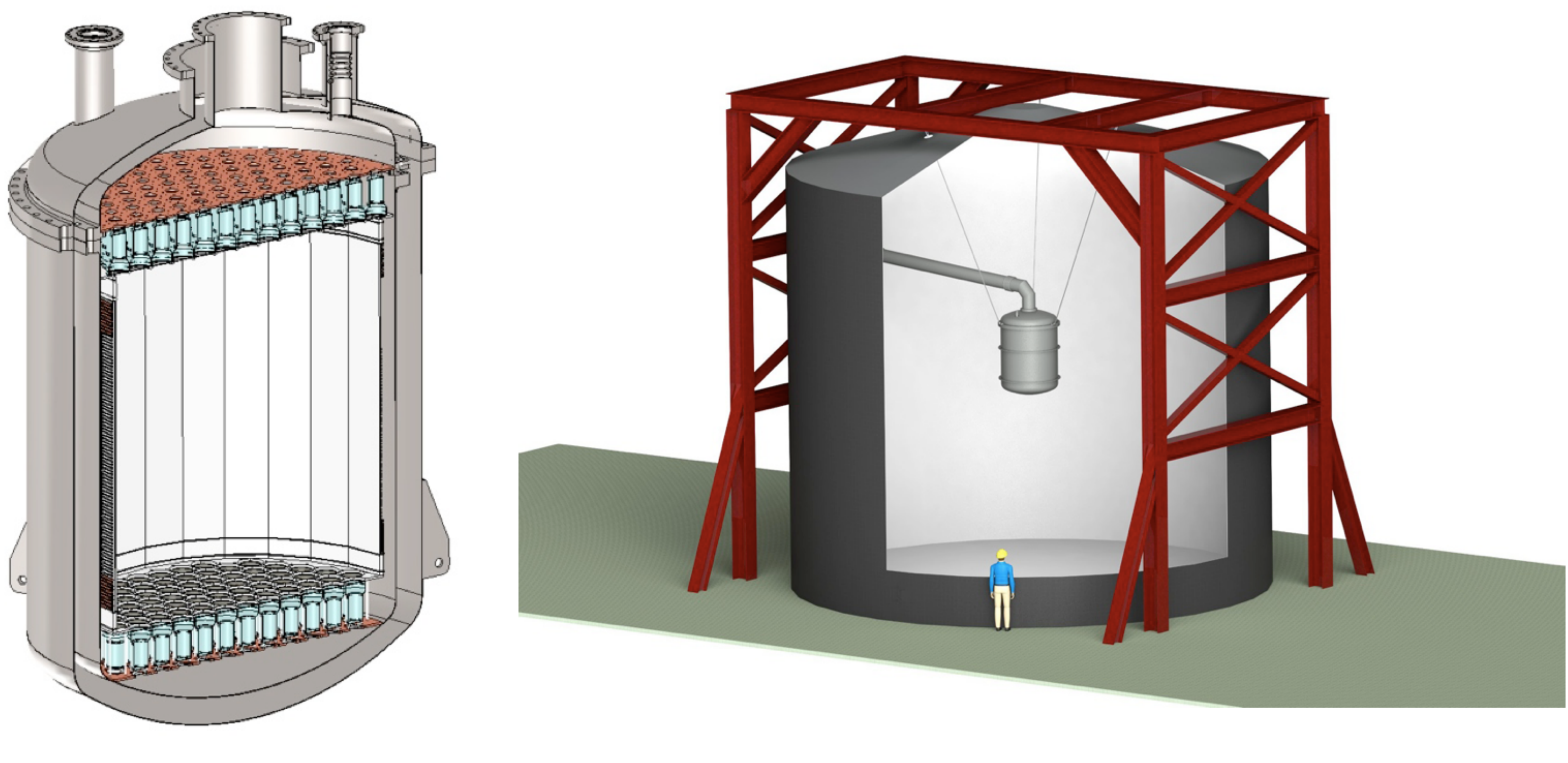}
\caption{{\bf Left:} Cross sectional drawing of the \Xeton\ cryostat containing the PTFE that bounds the active LXe, PMTs and support structures, field shaping rings and wire-mesh electrodes. 
{\bf Right:} The cryostat will be suspended at the center of an active water shield. Here the cryostat with only the central pipe is shown in the tank with the external support structure.}
\label{fig:cryostat}
\end{figure}

\section{Background and Expected Sensitivity}
\label{back_sens}

\Xeton\ will rely on a number of well established and proven techniques to achieve an unprecedented low background. First: the selection of every component used in the experiment will be based on an extensive radiation screening campaign, using a variety of complementary techniques and dedicated measurements as established for \Xehund~\cite{GATORscreening:2011,GEMPI}.  
Second: the self-shielding of LXe is exploited to attenuate and moderate radiation from material components within the TPC and simultaneously a fiducial volume will be defined, thanks to the TPC event imaging capability.  Additionally, the increased target mass provides powerful multiple scatter rejection, identifying background events. Third: radioactive elements within the LXe (such as from Kr or Rn  contamination) are reduced to a level which makes their contribution to the background negligible~\cite{Abe:2009,TAUP:2011}. Fourth: the \Xeton\ detector is  surrounded on all sides by a 4~m thick water shield, implemented as an active Cherenkov muon veto; this  shield is very effective in reducing the neutron and $\gamma$-ray background from the underground cavern rock, or from cosmic muon-induced events to negligible levels.  Finally, the TPC is designed to minimize light leakage from charge insensitive regions and events with rare topologies.

\begin{figure}[t]
    \includegraphics[width = .5\columnwidth]{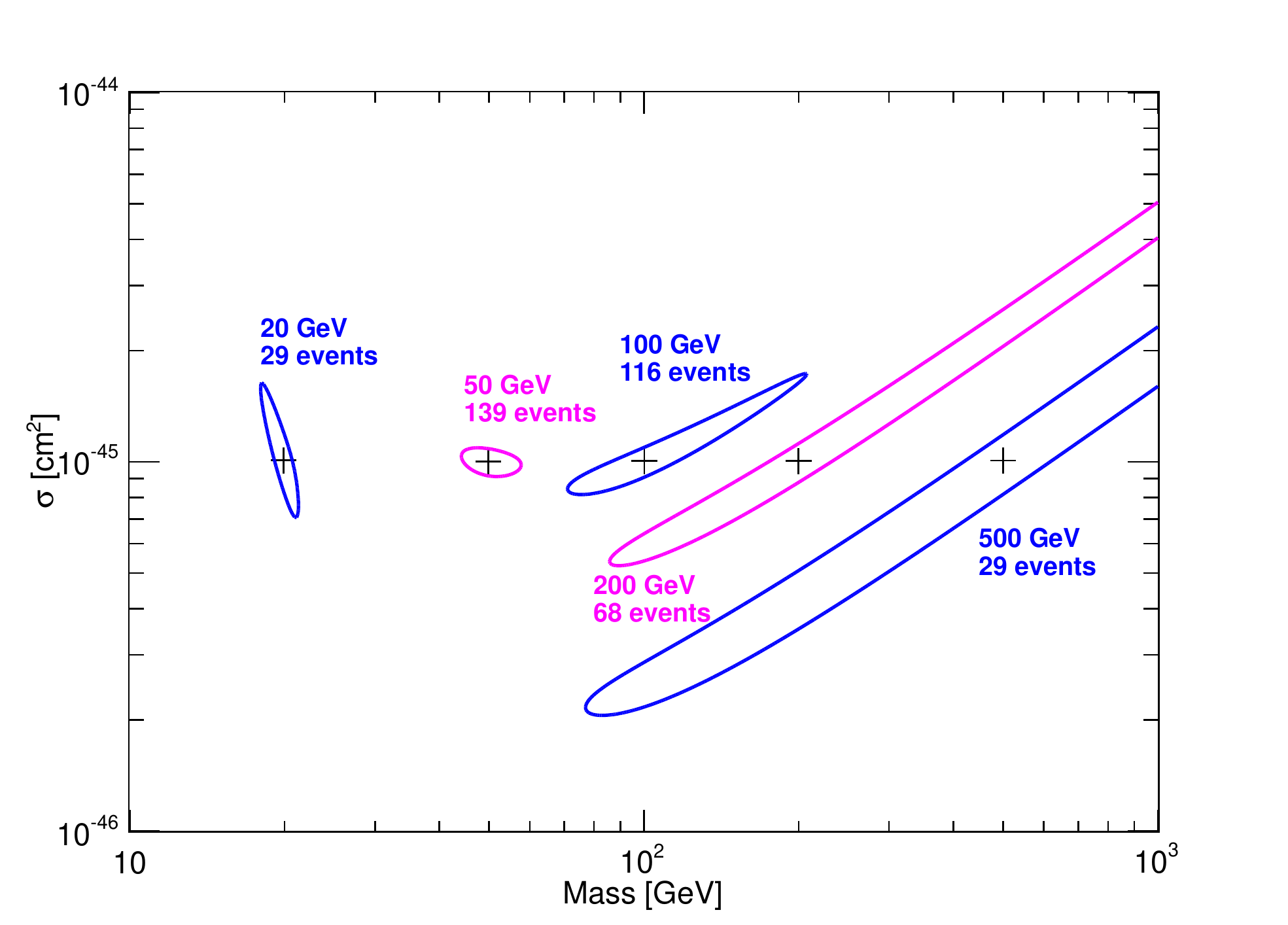}
    \includegraphics[width = .5\columnwidth]{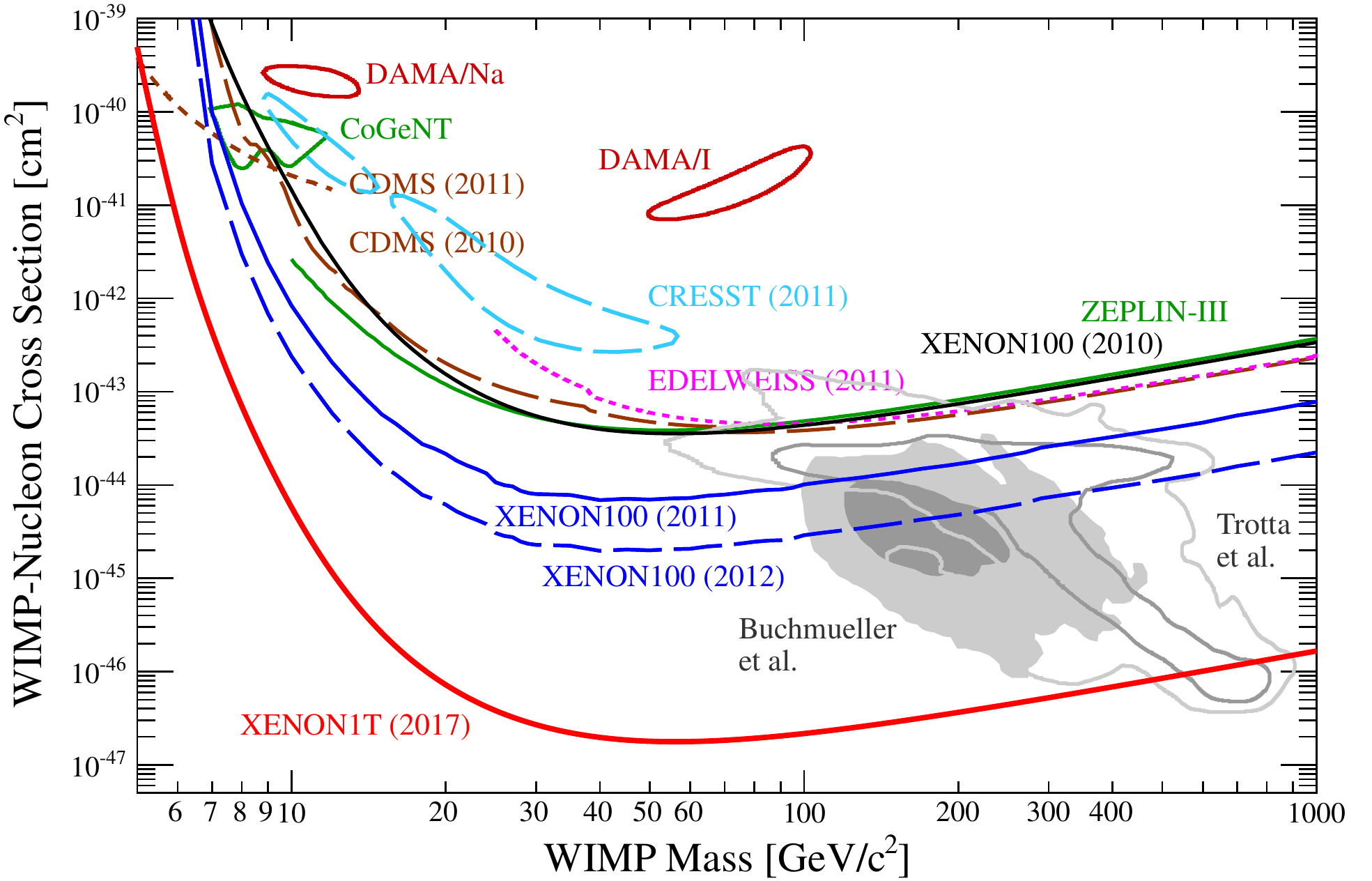}
    \caption{\small {\bf Left:} $1\sigma$ uncertainties in determining WIMP mass and $\sigma_{WIMP-N}$ from 2.2 ton-years with \Xeton. WIMP masses of 20, 50, 100, 200, 500 GeV/c$^{2}$ and $\sigma_{WIMP-N}$ of $10^{-45}\,\n{cm}^{2}$. {\bf Right:} Achieved and projected limits on $\sigma_{\n{SI}}$ from the \Xehund\ and \Xeton\ detectors. For comparison, results from a selection of other experiments are shown along with the most likely parameter space for a detection as predicted by the Constrained Minimal Supersymmetric Extension of the Standard Model~\cite{Trotta:2008,Buchmueller:2011}.}
   \label{fig:wimp_csmass_2}
\end{figure}

The experiment aims to reduce background from all expected sources such that the fiducial mass and the low energy threshold will allow \Xeton\ to reach an unprecedented sensitivity, ideally matched to probe a particularly rich region of electroweak scale parameter space, with a realistic WIMP discovery potential. With 2 years live-time and $1.1\,\n{ton}$ fiducial mass, \Xeton\ could detect on the order of 100 dark matter events, assuming $\sigma_{\n{SI}}$ $\sim$ $10^{-45}\,\n{cm}^2$ and for a WIMP mass of $100~\,\n{GeV/c^2}$.
With such a signal, \Xeton\ would be able to significantly constrain the WIMP cross-section and mass. Fig.~\ref{fig:wimp_csmass_2} (Left) show the $1\sigma$ uncertainties for the interaction cross section as a function of WIMP mass, assuming $\sigma_{\n{SI}}  = 10^{-45}\,\n{cm}^{2}$.
In the absence of a positive signal, the experiment aims to  exclude cross-sections above $\sigma_{\n{SI}}\sim
2\times 10^{-47} \,\n{cm}^2$ at $90\%\,\n{CL}$ for $50~\,\n{GeV/c^2}$ WIMPs, Fig.~\ref{fig:wimp_csmass_2} (Right), such that the bulk of the theoretically favored parameter space for SUSY WIMPs can be excluded~\cite{Trotta:2008,Buchmueller:2011}. 

\input{referenc}

\end{document}

%% file: referenc.tex
%
%
%